\documentclass[twocolumn]{aastex62}

\usepackage{graphicx}
\usepackage{dcolumn}
\usepackage{bm}
\usepackage{color}
\usepackage{xcolor}
\usepackage{amsmath}
\usepackage{lineno}

\newcommand{\xzy}[1]{{\color{red}  #1}}

\newcommand\lsim{\mathrel{\rlap{\lower4pt\hbox{\hskip1pt$\sim$}}
\raise1pt\hbox{$<$}}}
\newcommand\gsim{\mathrel{\rlap{\lower4pt\hbox{\hskip1pt$\sim$}}
\raise1pt\hbox{$>$}}}
\graphicspath{{./}{figures/}}

\revised{\today}
\submitjournal{ApJL}

\begin{document}


\title{Localizing Dynamically-Formed Black Hole Binaries in Milky Way Globular Clusters with LISA}


\correspondingauthor{Zeyuan Xuan}
\email{zeyuan.xuan@physics.ucla.edu}

\author{Zeyuan Xuan}
\affiliation{ Department of Physics and Astronomy, UCLA, Los Angeles, CA 90095}
\affiliation{Mani L. Bhaumik Institute for Theoretical Physics, Department of Physics and Astronomy, UCLA, Los Angeles, CA 90095, USA}

\author{Kyle Kremer}
\affiliation{Department of Astronomy and Astrophysics, University of California, San Diego, 9500 Gilman Drive, La Jolla, CA, 92093, USA}

\author{Smadar Naoz}
\affiliation{ Department of Physics and Astronomy, UCLA, Los Angeles, CA 90095}
\affiliation{Mani L. Bhaumik Institute for Theoretical Physics, Department of Physics and Astronomy, UCLA, Los Angeles, CA 90095, USA}

\begin{abstract}
{
The dynamical formation of binary black holes (BBHs) in globular clusters (GCs) may contribute significantly to the observed gravitational wave (GW) merger rate. Furthermore, LISA may detect many BBH sources from GCs at mHz frequencies, enabling the characterization of such systems within the Milky Way and nearby Universe. In this work, we use Monte Carlo N-body simulations to construct a realistic sample of Galactic clusters, thus estimating the population, detectability, and parameter measurement accuracy of BBHs hosted within them. In particular, we show that the GW signal from $0.7\pm 0.7$, $2.0\pm 1.7$, $3.6\pm 2.3$, $13.4\pm 4.7$ BBHs in Milky Way GCs can exceed the signal-to-noise ratio threshold of $\rm SNR =30$, 5, 3, and 1 for a 10-year LISA observation, with $\sim 50\%$ of detectable sources exhibiting high eccentricities ($e\gtrsim0.9$). Moreover, the Fisher matrix and Bayesian analyses of the GW signals indicate these systems typically feature highly-resolved orbital frequencies ($\delta f_{\rm orb}/ f_{\rm orb} \sim 10^{-7}-10^{-5}$) and eccentricities ($\delta e/ e \sim 10^{-3}-0.1$), as well as a measurable total mass when SNR exceeds $\sim20$. Notably, we show that high-SNR BBHs can be confidently localized to specific Milky Way GCs with a sky localization accuracy of $\delta \Omega \sim 1$~deg$^2$, and address the large uncertainties in their distance measurement ($\delta R \sim 0.3 - 20$~kpc). The detection and localization of even a single BBH in a Galactic GC would allow accurate tracking of its long-term orbital evolution, enable a direct test of the role of GCs in BBH formation, and provide a unique probe into the evolutionary history of Galactic clusters.
}
\end{abstract}


\section{Introduction} \label{sec:intro}

The LIGO/Virgo/KAGRA (LVK) collaboration \citep[e.g., ][]{2021arXiv211103634T} has detected approximately 100 extragalactic binary black holes (BBH) mergers \citep[e.g.,][]{Abbott_2023}. However, the formation channels for these mergers remain unclear, with various proposed mechanisms potentially contributing, including isolated binary evolution \citep[e.g.,][]{Belczynski16,Stevenson17,Eldridge19}; dynamical formation in galactic centers \citep[][]{Kocsis_2012,Hoang+19,Stephan+19,Arca+23,Zhang24}, galactic fields \citep[][]{Michaely+19,Michaely+20,Michaely+22,Stegmann_2024}, globular clusters (GCs) \citep[e.g.,][]{Rodriguez2016, Fragione19,Kremer_2020}, active galactic nuclei disks \citep{Tagawa+2021,peng2021,Samsing+2022,Munoz+22,Gautham+23}, hierarchical triple systems \citep{Wen03,Naoz16,Hoang+18,Antonini+19}, and primordial black hole scenarios \citep{Bird16,Sasaki16}. A key challenge in distinguishing these channels arises from the large distances and limited sky localization accuracy of extragalactic BBH mergers \citep[e.g.,][]{Vitale18}, which prevent confident identification of specific host environments. On the other hand, the future Laser Interferometer Space Antenna (LISA) \citep{2017arXiv170200786A} will observe BBHs in a lower frequency band ($10^{-4}-10^{-1}~\rm Hz$), which potentially allows us to probe their earlier evolutionary stages in the local Universe and distinguish between different hosting environments \citep[see, e.g., ][]{barack04,Mikoczi+12,robson18, chen19, Hoang+19, Fang19, tamanini19, breivik20,Torres-Orjuela21,Wang+21,Xuan+21,Zhang+21,Naoz+22,amaro+22,Naoz+23}. 

In particular, GCs are considered ideal places for the dynamical formation of mHz BBHs \citep[][]{miller02hamilton,PortegiesZwart+McMillan02,Morscher2015,rodriguez15,rodriguez16,samsingorazio18,Samsing+18,Orazio+18,Zevin_2019,Kremer2019_initialsize,gerosa19,Vitale19}. For example, a large number of black holes (BHs) can form through stellar evolution in clusters \citep[see, e.g., ][]{Kroupa2001,Morscher2015}. Due to mass segregation, these BHs gradually sink to the cluster core on sub-gigayear timescales, leading to the formation of a BH-dominated central region \citep{Spitzer1969, kulkarni93,sigurdsson93}. In the
BH-dominated core, dynamically hard BH binaries promptly form through three-body interactions and further harden through gravitational wave (GW) capture, binary-single, and binary-binary scattering events \citep{heggie2001,Merritt2004,Mackey2007, Breen2013, Peuten2016, Wang2016gc,Arca18, kremer18,Zocchi2019,Zevin_2019,Kremer_2020,Antonini_2020}. Collectively, these dynamical processes result in BBH merger events at an estimated rate of $\mathcal{R}_0 \approx 7.2_{-5.5}^{+21.5} \mathrm{Gpc}^{-3} \mathrm{yr}^{-1}$ in the local universe \citep[e.g.,][]{Rodriguez2016,Kremer_2020,Antonini_2020}, potentially making a significant contribution to the total GW merger rate detected by the LVK.


Moreover, dynamically-formed BBHs in GCs can provide valuable information about their astrophysical environment. For example, BBH mergers in a dense stellar environment typically have non-negligible eccentricity \citep[e.g.,][]{O'Leary+09,Thompson+11,Aarseth+12,Kocsis_2012,breivik16,Orazio+18,Zevin_2019,Samsing+19,Martinez+20,Antonini+19,Kremer_2020,wintergranic2023binary,rom2024dynamicssupermassiveblackholes}, which could greatly enhance our understanding of their formation mechanisms  \citep[see, e.g.,][]{east13, samsing14, Coughlin_2015,breivik16,vitale16,nishizawa16b,Zevin2017, Gondan_2018a, Gondan_2018b, Lower18, Romero_Shaw_2019, moore19,2021ApJ...913L...7A, 2021arXiv211103634T,Zevin_2021}, improve parameter estimation accuracy \citep{Xuan23acc}, or help with detecting the presence of tertiary companions through eccentricity oscillations \citep{Thompson+11,Antognini+14,Hoang+18,Stephan+19,Martinez+20,Hoang+20,Naoz+20,Stephan+19,Wang+21, Knee2022ecc}. Further, as shown by recent studies, many BBHs undergo a wide (semi-major axis $a\gtrsim 0.1 \,\rm au$), highly eccentric (eccentricity $e\gtrsim 0.9$) progenitor stage before the final merger \citep[see, e.g.,][]{Kocsis_2012,Hoang+19,Xuan+23b,knee2024detectinggravitationalwaveburstsblack}, which can also have unique imprints on mHz GW detections \citep[][]{Xuan24bkg,Xuan24parameter}.


In this work, we will explore the properties of dynamically-formed BBHs in Milky Way GCs, with a focus on their detectability and parameter measurement accuracy in the mHz GW detection of LISA. Particularly, previous studies have shown that LISA can detect a handful of BBHs formed through {\it isolated} binary channels in the Milky Way \citep[see, e.g.,][]{lamberts18,Sesana2019jmu,Wagg2022,Tang2024}. On the other hand, with approximately 150 GCs in the Milky Way \citep[e.g.,][]{Harris+96,Baumgardt+18}, we expect a significant number of {\it dynamically}-formed BBH sources from these clusters \citep[see, e.g.,][]{kremer18}, for which more specific predictions of their population are still needed.

We highlight that the detection of even a single BH in Milky Way GCs would provide a \textit{direct test} of the role of GCs in BBH formation, offering valuable insights into the relative contributions of different formation channels to the observed BBH population. Therefore, it is essential to create a realistic sample of globular clusters that harbor binary black holes in the Milky Way, which will enable us to constrain the underlying populations of black holes and BBHs in GCs and assess which specific Galactic globular clusters are most likely to host resolvable BBHs within the LISA frequency band.

This paper is organized as follows. In Section~\ref{sec:simulation}, we introduce the simulation of globular clusters (the \texttt{CMC Cluster Catalog} cluster model) used in this work. Next, we fit the simulated clusters to Galactic globular clusters, thus estimating the population of Galactic BBHs in GCs today. Based on the simulations, we estimate the number of BBHs detectable by LISA, their eccentricity and mass distributions (see Section~\ref{subsec:detectability}), and which specific Galactic GCs are most likely to host these sources (see Section~\ref{sec:location}). Furthermore, we adopt the Fisher matrix and Bayesian analyses in Section~\ref{sec:astroinfer}, and assess the parameter measurement accuracy of BBHs in the Milky Way GCs. In Section~\ref{sec:discussion}, we summarize the results and discuss the astrophysical implications. Throughout the paper, unless otherwise specified, we set $G=c=1$.

\section{Creating a mock Galactic sample}
\label{sec:simulation}

To assemble a realistic sample of globular cluster BBHs, we use the \texttt{CMC Cluster Catalog} cluster models of \citet{Kremer_2020}. This suite of models is computed using the Monte Carlo $N$-body dynamics code \texttt{CMC}, which includes the most up-to-date physics for studying the formation and evolution of BHs in dense clusters \citep[for review, see][]{Rodriguez2022}. Each of the 148 independent simulations of this catalog models the gravitational dynamics and stellar evolution for each of the $N$ stars in the system, keeping track of various properties (e.g., masses, semi-major axis, eccentricity) of all BBHs present in the cluster from formation to present day. The \texttt{CMC Cluster Catalog} has been tested rigorously against observations of Galactic globular clusters and successfully reproduces global features like surface brightness profiles, velocity dispersion profiles, and color-magnitude diagrams \citep{Rui2021} as well as specific compact object populations including millisecond pulsars \citep{Ye2019}, X-ray binaries \citep{Kremer2019_initialsize}, cataclysmic variables \citep{Kremer2021_wd}, and connections to BBHs observed by LIGO/Virgo \citep{Kremer_2020}.

For this study, we aim to predict which specific Galactic GCs are most likely to host resolvable LISA sources at present. In this case, we must identify a single ``best-fit'' model from the \texttt{CMC Cluster Catalog} for each observed Milky Way GC. The \texttt{CMC Catalog} models can be sorted into a three-dimensional grid in $M_{\rm cl}-Z-R_{\rm gc}$ space, where $M_{\rm cl}$ is the total cluster mass at present (set by our choice of initial $N=[2,4,8,16,32]\times10^5$ stars and evolved to the current value), $Z=[0.01,0.1,1]\times Z_{\odot}$ is the current cluster metallicity, and $R_{\rm gc}=[2,8,20]\,$kpc is the current Galactocentric position. For a given observed GC, we identify the appropriate $M_{\rm cl}-Z-R_{\rm gc}$ model bin based on that GC's observed mass and metallicity values \citep[taken from][]{Harris2010}. Each distinct $M_{\rm cl}-Z-R_{\rm gc}$ bin contains 4 models of varying initial virial radius, which determines the clusters' final core and half-light radius. To find the single best-fit model, we identify the model in the appropriate $M_{\rm cl}-Z-R_{\rm gc}$ bin which has core radius at $t=12\,$Gyr closest to the observed core radius value \citep{Harris2010}.\footnote{We also tried fitting using half-light radius in place of core radius, and found no major differences in our results.} Once the best fit model is identified, we select a cluster age by randomly sampling from that model's available time snapshot outputs in the range $8-13.7\,$Gyr (typically this consists of a sample of 10-50 snapshots per model). This range is intended to reflect the uncertainty in cluster ages at present. Once a cluster snapshot is selected, we then identify the number and properties of all BBHs present in the best-fit model at that time. By repeating these steps for each observed GC, we build a sample of the full population of BBHs present in each of the Milky Way's GCs at present. We then repeat this full procedure ten times to assemble ten separate Galactic realizations (the random time snapshot draw enables us to create distinct ten distinct samples).

Once this population is assembled, we compute the gravitational wave strain for each of the individual BBHs predicted within each Galactic cluster at present. Component masses, semi-major axes, and eccentricities are obtained  directly from our best-fit \texttt{CMC} snapshot and the source distance and position is simply the heliocentric distance of the GC of interest \citep{Harris2010}.

\begin{figure}
    \centering
    \includegraphics[width=3.5in]{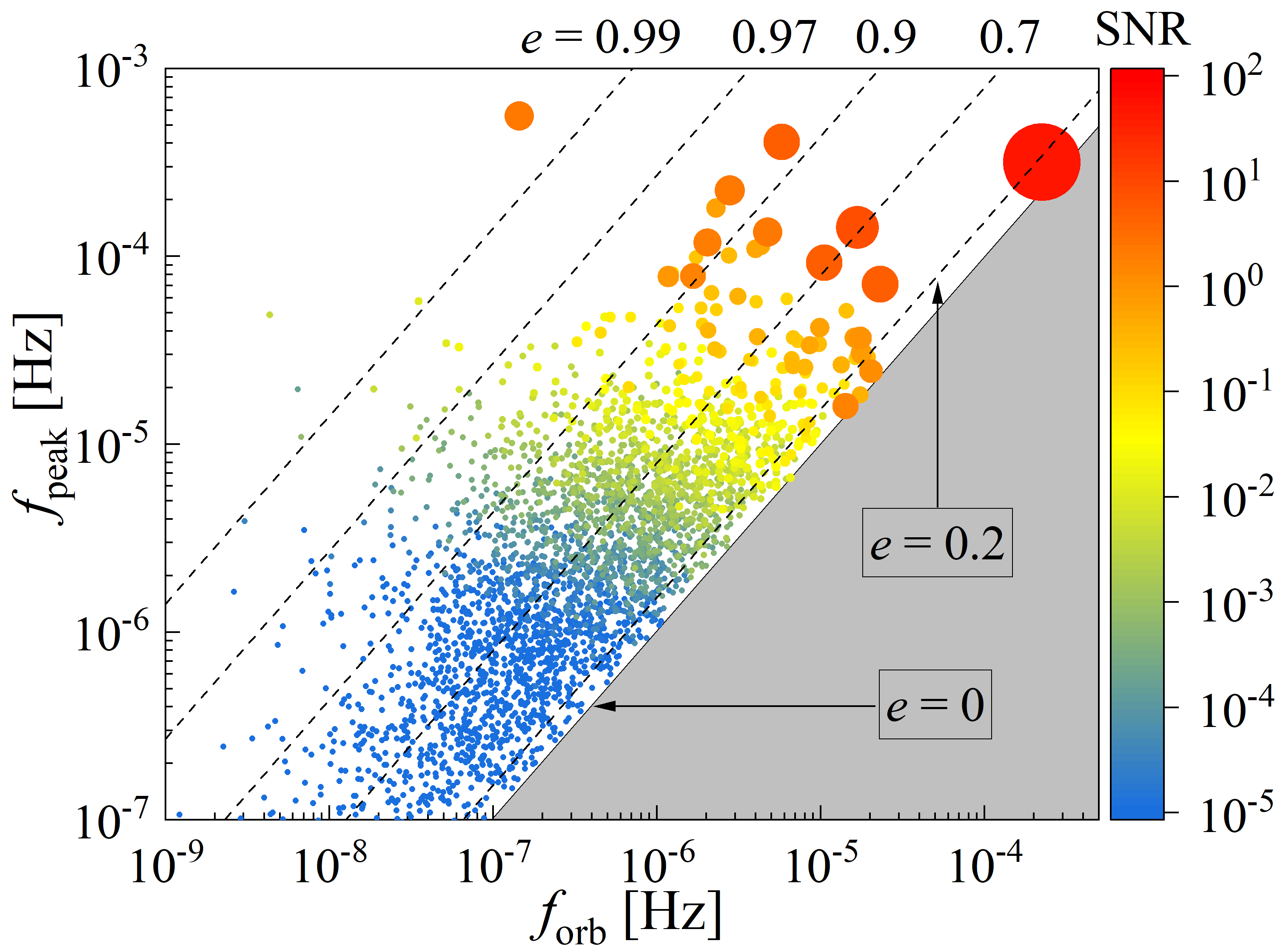}
    \caption{{\bf An example of simulated BBHs population in Milky Way GCs.} Here, we plot the orbital frequency and peak GW frequency of BBH systems from a single realization of our simulated Milky Way GCs (see Equation~(\ref{eq:peakf})). The SNR of each BBH, estimated analytically using Equation~(\ref{eq:snrsum}) for a 5-year LISA observation, is represented by different colors. Detectable BBHs, based on their SNR, are highlighted with enlarged dots. Dashed lines indicate the eccentricity of BBHs in different regions, with $e = 0.99, 0.97, 0.9, 0.7, 0.2, 0$, from left to right. The simulated BBHs have an average component mass of $m_1=15.56~{\rm M_{\odot}},\,m_2=15.21~{\rm M_{\odot}}$ for in-cluster BBHs and $m_1=20.73~{\rm M_{\odot}},\,m_2=20.46~{\rm M_{\odot}}$ for ejected BBHs, but the mass of individual systems may vary.} 
    \label{fig:population}
\end{figure}

Figure~\ref{fig:population} shows an example of the simulated BBH population
in Milky Way GCs. In particular, we choose one representative Galactic realization from the best-fit result of the \texttt{CMC Catalog}, and plot the orbital frequency ($f_{\rm orb}$) and the peak GW frequency ($f_{\rm peak}$) of each BBH system \citep[e.g.,][]{O'Leary+09}:
\begin{equation}
    f_{\rm peak} = f_{\rm orb} (1+e)^{1/2} (1-e)^{-3/2}\, .
    \label{eq:peakf}
\end{equation}

We note that most of the dynamically-formed systems have non-negligible eccentricity. In this case, $f_{\rm peak}$ of the eccentric GW signal indicates the frequency of the peak GW power, which typically needs to be within the mHz band for LISA to detect the source. Furthermore, in Figure~\ref{fig:population}, we can estimate the eccentricity of each system by comparing $f_{\rm peak}$ and $f_{\rm orb}$ (see the dashed lines with $e=0.99,0.97,0.9,0.7,0.2,0$, from left to right). For example, a system with $f_{\rm peak} \sim f_{\rm orb}$ should have moderate eccentricity, and a system with $f_{\rm peak} \gg f_{\rm orb}$ is highly eccentric.

In Figure~\ref{fig:population}, we use different colors to represent the signal-to-noise ratio (SNR) of BBHs, which is estimated analytically by summing the contributions from all the harmonics of their GW signal \citep[see, e.g., ][]{peters63,kocsis12levin,Xuan+23b}:
\begin{equation}
    {\rm SNR}^2=8h^2_0(a)\sum_n\frac{g(n,e)}{S_n(nf_{\rm orb})n^2}T_{\rm obs} \ ,
    \label{eq:snrsum}
\end{equation}
in which $n=1,2,3...$ represents the number of harmonics, $T_{\rm obs}$ is the observation time, and $h_0(a)=\sqrt{32/5}\,m_1 m_2/(R a)$ depends on the binary's component mass $m_1,\, m_2$, semi-major axis $a$, and distance $R$. Additionally, $S_{\mathrm{n}}(f)$ is the
spectral noise density of LISA evaluated at GW frequency $f$
\citep[we adopt the LISA-N2A5 noise model, see, e.g.,][]{2016PhRvD..93b4003K,Robson+19}, and $g(n,e)$ can be evaluated using:
\begin{eqnarray}
g(n, e) & =\frac{n^4}{32}\left[\left(J_{n-2}-2 e J_{n-1}+\frac{2}{n} J_n+2 e J_{n+1}-J_{n+2}\right)^2\right. \nonumber \\
& \left.+\left(1-e^2\right)\left(J_{n-2}-2 J_n+J_{n+2}\right)^2+\frac{4}{3 n^2} J_n^2\right],\label{eq:gne}
\end{eqnarray}
in which $J_i$
is the $i$-th Bessel function evaluated at $ne$. 

We note that the estimation of $\rm SNR$ in Equation~(\ref{eq:snrsum}) is based on the sky-average power of GW emission. In reality, the inclination and sky location of the GW source will also affect the detected $\rm SNR$, making it different from the average value. Therefore, the value of SNR shown in Figure~\ref{fig:population} should be understood as a heuristic estimation (in most cases, the variation in SNR caused by different inclinations is within an order of magnitude of the average value). For a detailed discussion, see the appendix of \citet{Xuan+23b}. 


As can be seen in Figure~\ref{fig:population}, the simulated orbital frequency and eccentricity of BBHs in Milky Way GCs distribute in a wide range of parameter space, with the majority of the binaries lying below the threshold of $\rm SNR=5$ for a 5-year LISA observation. However, there can be a handful of detectable sources with both moderate and high eccentricities. For example, we identified $3$ detectable sources in this realization, which have the orbital parameters of $f_{\rm orb} = 2.24\times 10^{-4}, 1.67\times 10^{-5}, 5.79\times 10^{-6}$~Hz ($a=0.0075, 0.048, 0.079$~au) and $e=0.17, 0.71, 0.93$, respectively. Furthermore, there is a larger number ($\sim 20$) of highly eccentric BBHs in the region of $\rm SNR\sim 1-5$ (see the orange and yellow dots), which could be detected by LISA given a longer observation time, or contribute to a stochastic background of GW bursts \citep[see, e.g., ][]{Xuan24bkg}.

\section{Source Localization and Astrophysical Implications}
\subsection{Detectability and Eccentricity Distribution}\label{subsec:detectability}
Based on the simulation in Section~\ref{sec:simulation}, we computed the expected number of detectable BBHs formed in Milky Way GCs. In total, we expect the GW signal from $0.7\pm 0.7$, $2.0\pm 1.7$, $3.6\pm 2.3$, $13.4\pm 4.7$ BBHs to exceed the threshold of $\rm SNR =30$, 5, 3, and 1, respectively, for a 10-year observation of LISA \footnote{Here the error bar reflects the standard derivative of BBH number in each SNR bin, accounting for 10 realizations in the simulation (see Section~\ref{sec:simulation}).}. Furthermore, the simulation yields significant eccentricity for all the detectable BBH systems, with the eccentricity ranges from 0.167 - 0.994 for systems with $\rm SNR> 3$, which is consistent with the expected eccentricity distribution of dynamically-formed binaries in GCs \citep[see, e.g.,][for a LISA source expectations from Newtonian modeling of GCs]{kremer18}. 

We highlight that $\sim 50\%$ of the BBHs with $\rm SNR>5$ have high eccentricity in the detection ($e\gtrsim0.9$); the fraction becomes even larger ($\sim 70\%$) for the population with $\rm SNR \sim 1-5$. This phenomenon reflects the highly eccentric nature of compact binary formation in a dense stellar environment. It also indicates that most of the GW signals from BBHs in Galactic GCs will be characterized by ``repeated bursts" \citep[][]{Xuan+23b}, for which the detectability and parameter extraction accuracy have been recently investigated \citep[][]{Xuan24parameter}. 

The large fraction of highly eccentric BBHs can be understood analytically. In particular, the GW signal from eccentric binaries is made up of multiple harmonics, some of which have frequencies much higher than the binary's orbital frequency (see, e.g., Equation~(\ref{eq:peakf})). Therefore, compared with circular BBHs, eccentric sources in GCs could enter the sensitive band of LISA with a much wider orbital separation $a$, when their $f_{\rm orb}$ is well below the mHz band. In other words, these eccentric binaries will be detected at the earlier evolution stages, with more extended lifetimes and larger number expectations than circular sources in the same frequency band. For example, highly eccentric, stellar mass BBHs can stay in the mHz GW band with a lifetime of \citep[][]{peters63,Xuan24bkg}:
\begin{equation}
    \begin{aligned}
    &\tau_{\mathrm{ecc}}\sim \frac{3}{85\mu M^2} a^4\left(1-e^2\right)^{7 / 2}\\ 
    &\sim  1.17\times10^{6}{\rm yr} \,\frac{2}{q(1+q)}\left(\frac{M}{20\rm M_{\odot}}\right)^{-3} \left(\frac{a}{1\rm au}\right)^{4}\left(\frac{1-e}{0.002}\right)^{\frac{7}{2}} ,
    \end{aligned}
    \label{eq:lifetime}
\end{equation}
where $M=m_1+m_2$, $\mu=m_{1}m_{2}/(m_1+m_2)$, and $q=m_1/m_2$. Note that this timescale is much longer than the merger timescale of a circular BBH system in mHz band (which typically lasts for $\sim 10^3-10^5$~years). Thus, these highly eccentric BBHs may dominate the population of BBHs in the local Universe, where their GW signals are strong enough to be identified. \footnote{However, stellar mass BBHs at larger distances are unlikely to be detected with high eccentricity (e.g., at a few hundred Mpc, detectable BBHs in GCs have eccentricities of at most 0.01, as shown in \citet{Kremer2019_LISA}). This is because highly eccentric sources, in general, have a smaller power of GW emission, but LISA can only detect high-SNR GW sources at cosmological distances.}

\begin{figure*}
    \centering
    \includegraphics[width=7in]{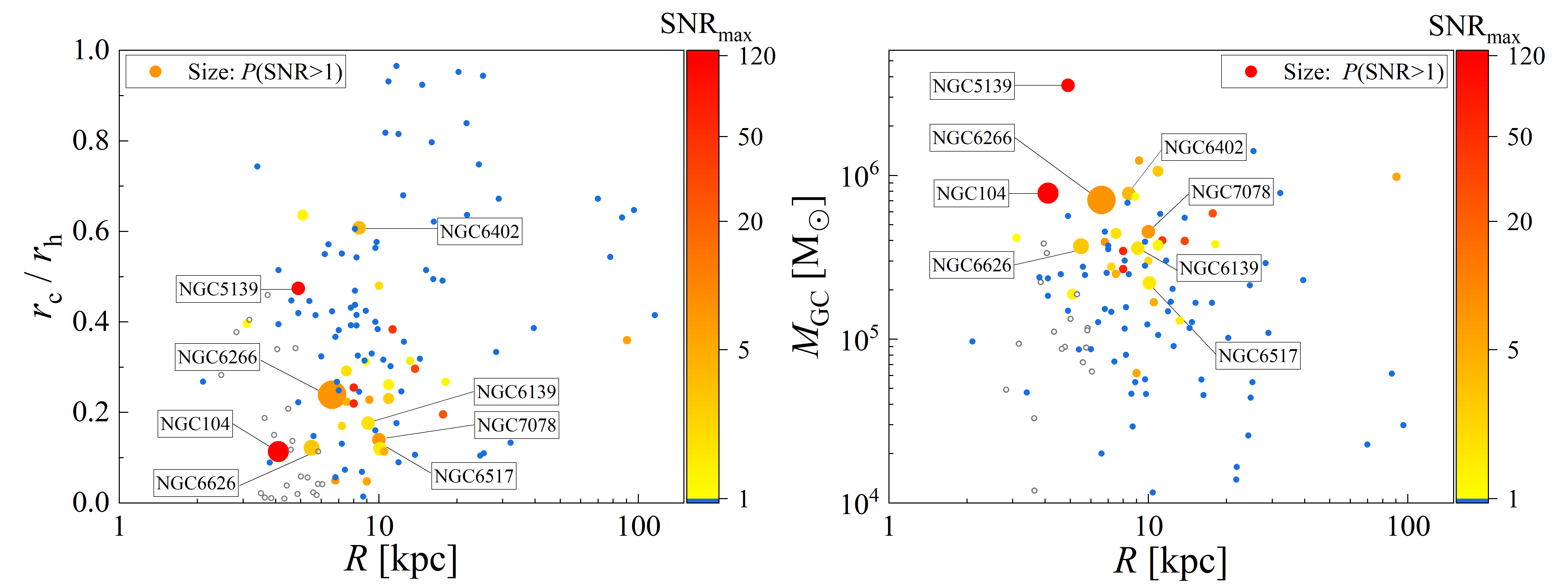}
    \caption{{\bf{Properties of Milky Way globular clusters likely to host detectable BBHs in the mHz GW band.}} Here we adopt the simulation results of the BBH population in the Milky Way GCs, based on 10 realizations of in-cluster sources. The size of solid dots represents the expected probability for a cluster to host a BBH system with $\rm SNR>1$, $P({\rm SNR>1})$, and the color represents the maximum $\rm SNR$ of detectable BBHs, assuming a 10-year observation. {\it Left Panel} plots the distance $R$ (x-axis) against the ratio of the core radius to the half-light radius, $r_c/r_h$ (y-axis); 
    {\it Right Panel} plots the distance $R$ (x-axis) against the estimated total mass $M_{\rm GC}$ of each globular cluster. The largest dot corresponds to the GC hosting 9 BBHs with $\rm SNR>1$ across the 10 realizations (i.e., an expected probability of $\sim 90 \%$), while the smallest dots represent GCs hosting 1 BBH system with $\rm SNR>1$ (expected probability $\sim 10 \%$). Solid blue dots represent GCs with no detectable BBHs (maximum $\rm SNR<1$), and hollow gray dots represent GCs without any BBHs in the simulation. We highlight the names of the top 8 GCs most likely to host detectable BBHs.}
    \label{fig:GCs}
\end{figure*}

Additionally, binary mergers originating from GCs can be categorized into different types, based on their evolution history. On top of the binaries merging within the GC after dynamical interactions (in-cluster mergers), a significant fraction of BBHs can undergo multiple hardening encounters before being ejected from the cluster, eventually merging in the galactic field \citep[ejected mergers; see, e.g.,][]{downing10,rodriguez16,Rodriguez18,Kremer2019_LISA}. In our simulations, we find that in-cluster BBHs contribute approximately 0.7, 1.5, 2.2, and 5.8 GW sources, while the ejected population contributes around 0, 0.5, 1.4, and 7.6 GW sources above the thresholds of $\rm SNR = 30$, 5, 3, and 1, respectively, for a 10-year LISA observation. 


\subsection{The Location of Gravitational Wave Sources}
\label{sec:location}

Next, we analyze which specific Galactic globular clusters are most likely to host resolvable BBHs, using the \texttt{CMC Catalog} from \citet{Kremer_2020}. In particular, we take the observational properties of Milky Way GCs following \citet{Harris2010} (see dots in Figure~\ref{fig:GCs}), and choose the best-fit model in the \texttt{CMC Catalog} that matches the mass, metallicity, Galactic position, and core radius of each cluster (see Section \ref{sec:simulation}). For each fitted cluster, we then compute the population properties of BHs and estimate their expectation of hosting a detectable BBH system. 

The results are summarized in Figure~\ref{fig:GCs}. Specifically, {\it Left Panel} depicts the ratio of the clusters' core radius to the half-light radius, $r_c/r_h$ (y-axis), versus their distance, $R$, from the detector (x-axis); {\it Right Panel} shows the estimated total mass of each cluster, $M_{\rm GC}$, against their distance $R$ (x-axis) (see \citet{Harris2010}). In Figure~\ref{fig:GCs}, the size of solid dots represents the expected probability for a cluster to host a BBH system with $\rm SNR>1$:
 \begin{equation}
     P({\rm SNR>1})= \frac{N_{{\rm SNR>1}}}{N_{\rm realization}} \ ,
 \end{equation}
 where $N_{\rm realization}=10$ represents the total number of realizations for Milky Way GCs (see Section~\ref{sec:simulation}), and $N_{{\rm SNR>1}}$ represents the total number of BBHs with $\rm SNR >1$ in a GC, summed across all the realizations. 
 
 The color of dots in Figure~\ref{fig:GCs} represents the detectability of their largest SNR GW source in the simulation, with colors ranging from yellow to red. Furthermore, we use solid blue dots to represent GCs with no detectable BBHs (maximum $\rm SNR<1$) and hollow gray dots to represent GCs without any BBHs in the simulation at all. 
 In the Figure, we highlight the names of the eight fitted GCs that we predict are most likely to host detectable BBHs.
 


As illustrated in Figure~\ref{fig:GCs}, GCs with a high probability of hosting detectable BBHs tend to cluster within specific regions of the parameter space. Notably, we predict detectable BBHs are most likely to be resolved in GCs that are close in distance ($R\lesssim 10$~kpc), exhibit a small $r_c/r_h$ ($\sim 0.1-0.3$, indicative of denser, more dynamically-active GCs, see {\it Left Panel}), and have a large total mass (see {\it Right Panel}). 


\subsection{Astrophysical Implication}
\label{sec:astroinfer}
In this section, we first adopt the Fisher matrix analysis to explore the astrophysical information that can be extracted from dynamically-formed BBHs in Milky Way GCs. This method is commonly used as a linearized estimation of the parameter measurement error in the high $\rm SNR$ limit \citep[see, e.g., ][]{Coe+09, Cutler+94}. For completeness, we briefly summarize the relevant equations and waveform model used in this work \citep[see our previous works][for similar applications]{Xuan23acc, Xuan24parameter}.

We begin by defining the noise-weighted inner product between two gravitational waveforms, $h_{1}(t)$ and $h_{2}(t)$, as follows:
\begin{equation}
\left\langle h_{1} \mid h_{2}\right\rangle=2 \int_{0}^{\infty} \frac{\tilde{h}_{1}(f) \tilde{h}_{2}^{*}(f)+\tilde{h}_{1}^{*}(f) \tilde{h}_{2}(f)}{S_{\mathrm{n}}(f)} \mathrm{d} f \ ,
\label{eq:innerproduct}
\end{equation}
where $\tilde{h}_l$ (with $l=1,2$) denotes the Fourier transform of the waveform, and the star represents the complex conjugate.

Representing the parameters of a GW source as a vector $\boldsymbol{\lambda}$, the GW waveform $h$ can be expressed as $h(t;\boldsymbol{\lambda})$. The Fisher matrix is then defined as:
\begin{equation}
    F_{ij} = \left\langle\left.\frac{\partial h(\boldsymbol{\lambda})}{\partial \lambda_i}\right\rvert\, \frac{\partial h(\boldsymbol{\lambda})}{\partial \lambda_j}\right\rangle.\label{eq:fisherdefinition}
\end{equation}
where $\lambda_i$ denotes the i-th parameter of the waveform.

Let $C$ denote the inverse of the Fisher matrix, $C = F^{-1}$. This matrix approximates the sample covariance matrix of the Bayesian posterior distribution for the parameters of the GW source. Using this, we estimate the error in parameter measurement as follows:
\begin{equation}
\delta \lambda_{i}= \sqrt{\left\langle\left(\Delta \lambda_{i}\right)^{2}\right\rangle}=\sqrt{C_{i i}}\ .\label{eq:delt_estimation}
\end{equation}

To evaluate Equations~(\ref{eq:innerproduct}) - (\ref{eq:delt_estimation}) numerically, we further compute the GW signal, $h(t)$, from eccentric BBHs. Specifically, we adopt the x-model \citep{Hinder+10} for the waveform generation, assuming that the binaries undergo isolated evolution during observation. The x-model is a time-domain, post-Newtonian (pN)-based waveform family, designed to capture all key features introduced by eccentricity in non-spinning binaries \citep{Huerta+14}. It has been validated against numerical relativity for equal mass BBHs with $e=0.1$, covering 21 cycles before the merger, and also aligns with well-established waveform template families used in GW data analysis for the zero-eccentricity case \citep{Brown+10}. In this model, the binary orbit is described using the Keplerian parameterization at 3pN order, with the conservative evolution also given to 3pN order. The energy and angular momentum losses are mapped to changes in the orbital eccentricity $e$ and the pN expansion parameter $x \equiv (\omega M)^{2/3}$, where $\omega$ is the mean Keplerian orbital frequency. These two parameters evolve according to 2~pN equations. We note that, stellar-mass BBHs in the local universe typically have a pericenter distance larger than $\sim 10^{-3}{\rm au}$ in the mHz GW band (including the highly eccentric BBHs, see, e.g., \citet{Xuan+23b,Xuan24bkg}). Thus, their gravitational field is much weaker than the strength of the field for which the x-model has been validated against numerical relativity, and the x-model represents a plausible description of their GW signal\xzy{\footnote{Also, there have been recent studies focusing on fast and accurate waveform generation, such as for the case of eccentric extreme mass ratio inspirals (EMRIs) \citep[][]{Chua2021EMRI,Hughes2021EMRI,Katz2021EMRI}. However, the mass ($\sim 10\,{\rm M_{\odot}}$) and eccentricity range ($\sim 0.1-0.999$) of BBHs we discuss here are different. Therefore, we adopt the x-model for simplicity.}}.

\begin{figure*}[htbp]
    \centering
    \includegraphics[width=7in]{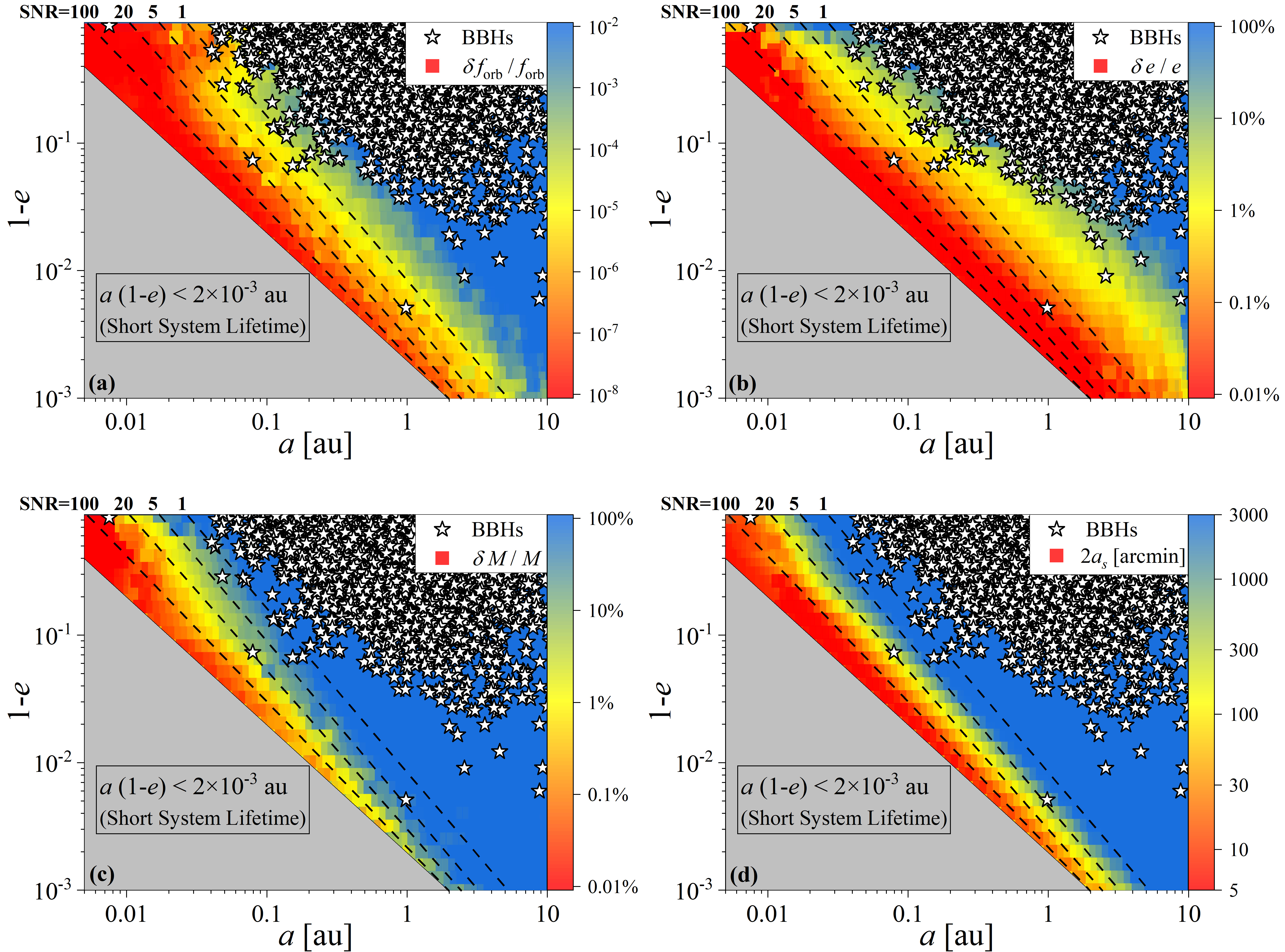}
    \caption{{\bf The population of BBHs from simulated Milky Way GCs, and their estimated parameter measurement error as a function of semi-major axis and eccentricity (for 5-yr observation). } Here, we adopt the simulation result of the compact binary population in the Milky Way globular clusters (see Section~\ref{sec:simulation}, note that here we show a single realization of the in-cluster and ejected BBHs from Milky Way GCs), and plot the semi-major axis and eccentricity (as $1-e$) of each BBH system (see the white stars). The background color maps the compact binary's parameter measurement error, for the orbital frequency ($\delta f_{\rm orb}/ f_{\rm orb}$, {\it Panel a}), eccentricity ($\delta e/ e$, {\it Panel b}), total mass ($\delta M/ M$, {\it Panel c}), and sky location (major axis of sky error ellipsoid, $2a_s$, {\it Panel d}). We note that, in reality, the BBHs can have different mass, inclination, and sky locations, which affects their parameter measurement. However, for simplicity, here we assume a fixed intrinsic parameter of $m_{1}=10$~M$_{\odot},$ $m_{2}=15$~M$_{\odot},$ $R=8$~kpc and $\Phi=\Theta=\theta=\phi=\psi=\pi/4$ when estimating the Fisher matrix and mapping the parameter measurement accuracy. 
    In each panel, the dashed lines represent equal signal-to-noise ratio contours (analytically calculated using Equation~(\ref{eq:snrsum})), with $\rm SNR=1,\, 5,\, 20,\, 100$ from right to left. We exclude the parameter space where the binary has a pericenter distance smaller than $a(1-e) < 2\times 10^{-3}\,\rm au$, since they have a short merger timescale and negligible number expectation in our simulation.}
    \label{fig:mapfisher2}
\end{figure*}

Furthermore, we include the detector's annual motion around the Sun to analyze the realistic detection of Milky Way BBHs (the detector response function, see, e.g., \citet{Cutler+98,Cornish+03,Kocsis2007} for more details). Consequently, the GW signal from an eccentric binary can be parameterized using:
\begin{equation}
h(t)=h(t;\boldsymbol{\lambda}=\left\{f_{\rm orb,0}, 1-e_0, M, q, \cos\Theta, \Phi, \cos\theta,\phi, R,\psi \right\})  , \label{eq:waveformparameter}
\end{equation}
in which $f_{\rm orb,0}, e_0$ are the initial orbital frequency and eccentricity of the binary \footnote{Hereafter, we use $f_{\rm orb}$ as an abbreviation for the initial orbital frequency $f_{\rm orb,0}$, which is related to the initial semi-major axis $a_0$ via $f_{\rm orb,0}=(2\pi)^{-1}M^{1/2}a_0^{-3/2}$; and use e as an abbreviation for the initial eccentricity $e_0$.}, $\Theta$ and $\Phi$ represent the spherical polar angles of the observer as viewed in the non-rotating, comoving frame of the compact object binary (i.e., the propagation direction of the GW signal viewed in the source’s frame); $\theta$ and $\phi$ are the spherical polar angles describing the sky location of the GW source viewed in the comoving frame of the solar system, where the LISA detector undergoes annual motion around the sun; $R$ is the binary's distance from the detector (which is set as the distance of the GC hosting this binary), and $\psi$ is the polarization angle of the GW signal. For more details on the parameterization, see Sec.~B in \citet{Xuan24parameter}.

After generating the GW signal as described by Equation~(\ref{eq:waveformparameter}),  we compute the partial derivatives of the waveform with respect to each parameter (see Equation~(\ref{eq:fisherdefinition})). For example, to calculate $\partial h/\partial M$, we vary the total mass $M\rightarrow M^{\prime}=M+\Delta M$, and generate a new waveform $h^{\prime}(t)=h^{\prime}(t;\boldsymbol{\lambda}^{\prime})$, where $\boldsymbol{\lambda}^{\prime}=\left\{f_{\rm orb}, 1-e, M+\Delta M, q, \cos\Theta, \Phi, \cos\theta,\phi, R,\psi \right\}$. The partial derivative is then approximated as $\partial h/\partial M \approx [h'(t) - h(t)]/\Delta M$. Note that each partial derivative is a time series representing the difference in the waveform caused by slightly varying one of the parameters around the central value. Also, in the numerical computation, we choose the parameter variation $d\lambda_i$ such that $\left\langle\left.dh\right\rvert\, dh\right\rangle \sim 10^{-3} \left\langle\left.h\right\rvert\, h\right\rangle$, ensuring the change in the waveform is small when computing the numerical derivative.

\begin{figure}[htbp]
    \centering
    \includegraphics[width=3.5in]{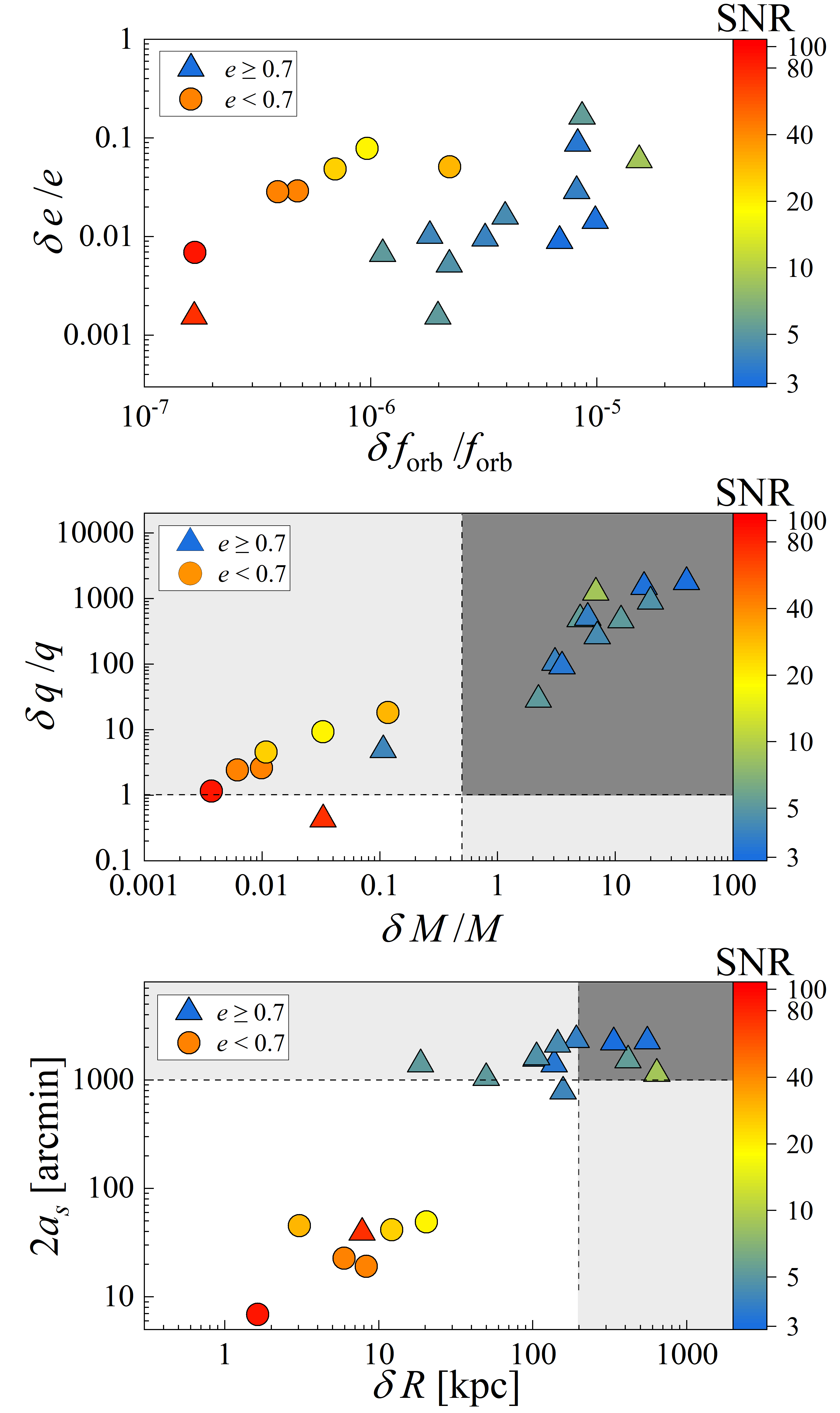}
    \caption{{\bf Parameter Measurement error of the detectable BBHs in simulated Milky Way GCs, estimated using Fisher matrix analysis.} Here we adopt the simulation results of BBHs population in Milky Way GCs, with a total of 10 realizations. Each point in the figure represents a BBH system in the simulation, with the color showing the $\rm SNR$ of its GW signal (assuming a 5-year observation) and the shape showing the eccentricity of the orbit ($e>0.7$, plotted in triangles; $e<0.7$, plotted in circles). {\it Upper Panel} shows the relative error of orbital frequency and eccentricity measurement. {\it Middle Panel} shows the relative error of total mass and mass ratio measurement. In {\it Bottom Panel}, we show the absolute error of the BBHs' distance measurement (x-axis), and their sky localization accuracy (as the major axis of the sky error ellipsoids, $2a_s$, on the y-axis). We exclude the parameter space where measurement accuracy is insufficient to help with the astrophysical interpretation, as marked using grey-shaded regions. }
    \label{fig:errorsum}
\end{figure}

Finally, we compute the inner products between derivative waveforms using Equation~(\ref{eq:innerproduct}), then substitute the results into Equation~(\ref{eq:fisherdefinition}) to calculate $F_{ij}$. The parameter measurement errors are then estimated by inverting $F_{ij}$, as described in Equation~(\ref{eq:delt_estimation})).  We summarize the results in Figures~\ref{fig:mapfisher2} - \ref{fig:errorsum} (note that Fisher matrix analysis yields the measurement error for all the parameters, $\left\{f_{\rm orb,0}, 1-e_0, M, q, \cos\Theta, \Phi, \cos\theta,\phi, R,\psi \right\}$, but here we only show some of them to avoid clustering).

In Figure~\ref{fig:mapfisher2}, we present a realistic example of the parameter measurement error for BBHs in Milky Way GCs. The white stars in the figure represent all the in-cluster and ejected BBHs from one realization of our simulated Milky Way GCs. In the background, we use different colors to show the parameter measurement errors,  $\delta \lambda_i$, as a function of the BBHs' orbital parameters, $(a,1-e)$. Particularly, {\it Panel a} shows the relative error in orbital frequency measurement, $\delta f_{\rm orb}/ f_{\rm orb}$; {\it Panel b} shows the relative error in eccentricity measurement, $\delta e/ e$; {\it Panel c} shows the relative error in total mass, $\delta M/ M$; and {\it Panel d} shows the absolute error in sky location ($2a_s$, as the major axis of the sky error ellipsoids, see e.g., \citet{Lang2006,Kocsis_2008,Kocsis2007,Mikoczi+12}). To ensure consistency in mapping the parameter measurement errors, we fix specific BBH parameters when computing the Fisher matrix results shown in Figure~\ref{fig:mapfisher2} ($m_{1}=10$~M$_{\odot},$ $m_{2}=15$~M$_{\odot},$ $R=8$~kpc and $\Phi=\Theta=\theta=\phi=\pi/4$, assuming a 5-year observation). However, BBHs in Milky Way GCs can have different mass, inclination, and sky locations, which potentially affects their parameter measurement \footnote{We note that, the parameter measurement accuracy in Figure~\ref{fig:mapfisher2} can be rescaled for BBHs with different distances, see Eq.14 in \citet{Xuan24parameter} for more details.}. Therefore, the value shown by background colors in Figure~\ref{fig:mapfisher2} should be interpreted as heuristic estimates of the realistic accuracy.

We highlight that the estimation shown in Figure~\ref{fig:mapfisher2} is agnostic to different BBH formation channels. In other words, for any potential BBH population in the galaxy, not necessarily from the GCs, their orbital parameter $(a,1-e)$ can be over-plotted on the figure to estimate the parameter measurement errors in a similar way. Moreover, Figure~\ref{fig:mapfisher2} indicates that our analysis of the parameter measurement accuracy is robust to variations in simulation results. For example, in {\it Panels $a$ and $b$}, regions with $\rm SNR>1$ (to the left of the dashed line at $\rm SNR=1$) show $\delta f_{\rm orb}/ f_{\rm orb}\lesssim 10^{-5}$ and $\delta e/e\lesssim 10\%$. This suggests that, in realistic observations, BBH systems detected in the Milky Way GCs can typically achieve high accuracy measurements of $f_{\rm orb}$ and $e$. Furthermore, as shown by {\it Panel c}, BBHs with $\rm SNR>20$ can have a mass measurement accuracy of $\delta M/M\lesssim 10\%$. However, marginally detectable sources may have poorly constrained total mass (blue regions near the dashed line of $\rm SNR =5$). Similarly, {\it Panel d} shows that BBHs with $\rm SNR>20$ can have a sky localization accuracy of $2a_s\lesssim 300$~arcmins (yellow regions), which indicates that high-SNR BBHs can generally be localized with an angular resolution of a few degrees in the Milky Way.







In Figure~\ref{fig:errorsum}, we summarize all BBHs with $\rm SNR>3$ (assuming a 5-year observation) from the simulation and compute their exact parameter measurement errors. These BBHs are drawn from 10 realizations in total, excluding the ejected population due to the poor constraints on their location within the Milky Way. In particular, {\it Upper Panel} shows the relative errors in orbital frequency and eccentricity measurements; {\it Middle Panel} shows the relative error in total mass and mass ratio measurements; and {\it Bottom Panel} shows the absolute error in the distance (x-axis) and sky localization (y-axis) measurements. In each panel, we use different colors to show the SNR of BBHs and the shape of dots to show the eccentricity ($e > 0.7$, plotted in triangles; $e < 0.7$, plotted in circles). Additionally, we exclude the parameter space where measurement accuracy is insufficient for astrophysical interpretation, as marked using grey-shaded regions. Specifically, in the {\it Middle Panel}, regions with $\delta M/ M>50\%$ and $\delta q/ q>100\%$ are excluded, as the BBHs in these regions are indistinguishable from other compact binary sources, such as binary neutron stars (BNSs) or double white dwarfs (DWDs). Similarly, in the {\it Bottom Panel}, BBHs with $2a_s >1000$~arcmins ($\sim 5-10$ times the typical tidal radius of the Milky Way GCs) and $\delta R>200$~kpc (halo radius of the Milky Way) cannot be confidently localized within a Milky Way globular cluster.

As shown in the {\it Upper Panel}, most detectable BBHs in Milky Way GCs have well-constrained orbital frequency and eccentricity. In particular, the fractional error in orbital frequency, $\delta f_{\rm orb}/ f_{\rm orb}$, reaches an accuracy of $\sim 10^{-7}-10^{-5}$ for BBHs with $\rm SNR>3$, which indicates a frequency resolution of $\delta f_{\rm orb}\lesssim 10^{-10}$~Hz in LISA data analysis \citep[see, e.g., Eq.18 in][for an analytical explanation]{Xuan24parameter}. Furthermore, all sources in Figure~\ref{fig:errorsum} have eccentricity measurement errors below $\delta e/ e<1$, which allows us to confidently detect non-zero eccentricities and distinguish dynamically-formed BBHs from circular binaries created via isolated evolutionary channels. Additionally, highly eccentric sources typically exhibit higher accuracy in eccentricity measurement, with $\delta e/ e$ reaching $\sim 0.1\%-10\%$ for BBHs with $e>0.7$. This trend is consistent with the results of our previous works \citep[][]{Xuan23acc}, which shows that eccentricity can break the degeneracy of waveform and significantly enhance the parameter measurement. In general, accurate measurements of orbital frequency and eccentricity can provide valuable insights into the long-term evolution of BBHs, enable the detection of potential environmental effects, and help infer different formation channels.

\begin{figure*}[htbp]
    \centering
    \includegraphics[width=7in]{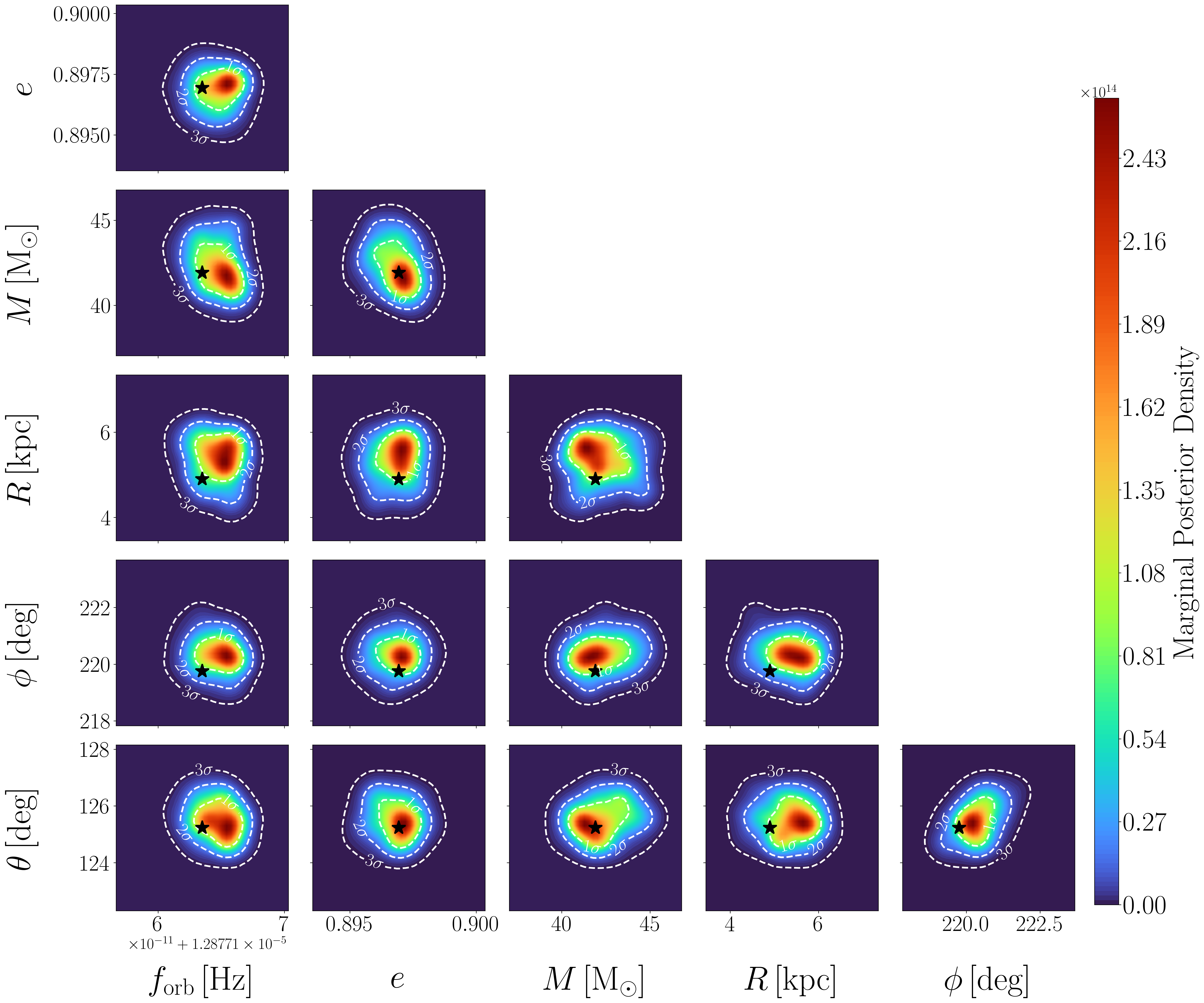}
    \caption{{\bf Parameter Measurement error of an example BBH system in Milky Way Globular Clusters, estimated using Bayesian analysis (for 5-yr observation).} Here we consider a BBH system with $m_{1}=23.3$~M$_{\odot},$ $m_{2}=18.6$~M$_{\odot},$ $f_{\rm orb}=1.2877164\times10^{-5}$~Hz, $e=0.897,$
    $R=4.9$~kpc, $\Phi=158.7^{\circ},\,\Theta=141.4^{\circ},\,\phi=219.9^{\circ},\,\theta=125.2^{\circ},\,\psi=153.9^{\circ}$. (Note that this system has $\rm SNR \sim 77$, and we keep 8 significant digits of $f_{\rm orb}$ to highlight its realistic measurement accuracy.) The color maps show the marginal posterior density of the binary's observed parameters, assuming a flat, 10-dimensional hyperrectangular prior and stationary Gaussian LISA noise. We use the star in each panel to indicate the system’s intrinsic parameters, and plot 68\%, 95\%, and 99.7\% credible regions for the marginal posterior densities (i.e, $1-3\,\sigma$ level sets defined analogously to a bivariate Gaussian density). To avoid clutter, we plot the results for the binary's intrinsic parameters $f_{\rm orb},\, e,\,M$; distance $R$; and sky location $\phi,\,\theta$, which are mostly relevant to its astrophysical inference. Additionally, the peak of posterior density exhibits a small offset from some parameters' intrinsic value (typically within $1\sigma$ level), which is caused by the random fluctuation of time-domain LISA noise and the limit of numerical accuracy for highly eccentric binary waveforms.
    }
    \label{fig:bayesian}
\end{figure*}

On the other hand, mHz GW detection may be less sensitive to the mass of BBHs in the Milky Way, primarily due to degeneracies in the waveform \footnote{Note that we expect most detectable BBHs in the Milky Way to be long-living systems, thus their orbital evolution is slow and their chirp rate is hard to measure \citep[see, e.g. Section 3.2 in][]{Xuan+23b}.}. For example, most BBHs in the {\it Middle Panel} do not have a well-constrained mass ratio, with the relative error of measurement exceeding $\delta q/ q \gtrsim 100\%$ (except for one highly eccentric BBH system with $\rm SNR\sim 75$, indicated by the red triangle in the bottom-left corner of {\it Middle Panel}). However, the total mass measurement is sufficiently accurate for a significant fraction of detectable BBHs, reaching $\delta M/M \sim 1\%-10\%$ for sources with $\rm SNR\gtrsim 20$ (see the red and orange dots to the left of the dashed line). Therefore, we expect LISA to accurately measure the mass of these high-SNR BBHs in the Milky Way GCs, making them distinguishable from BNSs and DWDs.

Furthermore, we highlight that, LISA can accurately localize most BBHs with $\rm SNR>20$ in Milky Way GCs (see the red and orange dots in {\it Bottom Panel}). In particular, these GW sources have a sky localization accuracy of $\sim 10-100$~arcmins (i.e., less than a few degrees), which is comparable to the typical diameter of Milky Way GCs ($\sim 10-120$~arcmins \footnote{Note that detectable BBHs are more likely to be hosted by nearby, massive GCs (see Section~\ref{sec:location}); as a result, their host clusters tend to have larger angular sizes than the general Milky Way GC population. For BBHs with $\rm SNR>1$, the average host GC diameter is $\sim 38$~arcmins (see Figure~\ref{fig:clustersize} in the Appendix).
}). Additionally, while the distance of BBHs is less precisely constrained compared to their sky location, most systems with $\rm SNR>20$ have a distance measurement error of $\delta R\sim 1 - 100$~kpc in the Milky Way. Given the significant distance to the nearby galaxies (e.g., the Andromeda Galaxy is located at $\sim 780$~kpc), we expect these BBHs to be distinguishable from extragalactic GW sources.

\subsection{Further validation of BBH localization using Bayesian Analysis}
\label{sec:bayesian}
As shown by Section~\ref{sec:astroinfer}, LISA has the potential to confidently localize a BBH system in specific Milky Way GCs, which could provide unique information about the Milky Way's compact binary formation and the dynamical environment of GCs. In this section, we further verify this point using a Bayesian analysis and address the uncertainty introduced by the BBHs' distance measurement.

First, we note that, although the Fisher matrix analysis adopted in Section~\ref{sec:astroinfer} has been widely used to estimate the parameter measurement error for LISA sources, this method can sometimes yield inaccurate results, particularly for degenerate parameters or parameters with weak influences on the shape of GW signal \citep[see, e.g.,][]{Vallisneri_2008, Toubiana_2020}. Consequently, the parameter measurement accuracy presented in Figures~ \ref{fig:mapfisher2} and \ref{fig:errorsum} should be interpreted as a heuristic estimation. Nevertheless, these estimates provide compelling evidence that BBHs in Milky Way GCs are promising targets for inferring their astrophysical properties.

To further validate the Fisher matrix results, we performed a full 10-dimensional Bayesian analysis for individual systems \citep[see, e.g.,][]{Finn92,Cutler+94,Christensen1998, Christensen01}, as shown in Figure~\ref{fig:bayesian} and Table~\ref{tab:compaison}. Particularly, in Figure~\ref{fig:bayesian}, we choose an example BBH system from the simulation results ($m_{1}=23.3$~M$_{\odot},$ $m_{2}=18.6$~M$_{\odot},$ $a=0.06332$~au, $e=0.897,$ $R=4.9$~kpc), and inject its GW signal $s_0(t;\boldsymbol{\lambda})$\footnote{Note that this system has $\rm SNR\sim77$ for 5-yr observation, and the detector response function is included in $s_0(t)$, see Section~\ref{sec:astroinfer}.} into a simulated stationary Gaussian LISA noise $n(t)$, which yields the mock LISA signal:
\begin{equation}
    z(t)=s_0(t;\boldsymbol{\lambda}) + n(t)\ .
    \label{eq:mocksignal}
\end{equation}

We then adopted Monte-Carlo sampling, generating GW templates $s(t;\boldsymbol{\lambda}')$ with parameters slightly different from their intrinsic values of $\boldsymbol{\lambda}$, and computing their inner product with the mock signal (see Equation~(\ref{eq:innerproduct})):
\begin{equation}
\left\langle z \mid s\right\rangle=2 \int_{0}^{\infty} \frac{\tilde{z}(f) \tilde{s}^{*}(f)+\tilde{z}^{*}(f) \tilde{s}(f)}{S_{\mathrm{n}}(f)} \mathrm{d} f \ .
\label{eq:innerproduct2}
\end{equation}

Assuming a uniform hyperrectangular prior, the posterior probability density of the observed parameters can be computed using the aforementioned inner products (see, e.g., Eqs 1-6 in \citet{Christensen01}):

\begin{equation}
    p(\boldsymbol{\lambda}'\mid z) \propto p(z\mid \boldsymbol{\lambda}') =K \exp [2\langle z\mid s(\boldsymbol{\lambda}')\rangle-\langle s(\boldsymbol{\lambda}')\mid s(\boldsymbol{\lambda}')\rangle]\ ,
    \label{eq:posterior}
\end{equation}
where K is a constant.

As shown in Figure~\ref{fig:bayesian}, the marginal posterior density of the binary’s observed parameters, as indicated by the color maps, yields a parameter measurement error comparable with the Fisher matrix estimation. Particularly, the Fisher matrix estimation of this example BBH system is represented by the red triangle to the bottom-left of each panel in Figure~\ref{fig:errorsum}, with its parameter measurement error $\delta f_{\rm orb}\sim2.1\times10^{-12}$~Hz, $\delta e\sim1.4\times 10^{-3},$ $\delta M\sim1.4$~M$_{\odot},$  $\delta R\sim7.8$~kpc, and sky localization accuracy $2a_s\sim39$~arcmin$\sim 0.65$~degrees. These values are generally consistent with the 1$\sigma$ levels in Bayesian analysis, which yields $\delta f_{\rm orb}\sim2.8\times10^{-12}$~Hz, $\delta e\sim1.2\times 10^{-3},$ $\delta M\sim3.1$~M$_{\odot},$  $\delta R\sim1.1$~kpc, and sky localization accuracy $\sim 1.3$~degrees (see Figure~\ref{fig:bayesian}). Therefore, the Bayesian study of this specific case partly justifies the Fisher matrix results of Figures~\ref{fig:mapfisher2} and \ref{fig:errorsum}. We also show the Bayesian analysis results of a BBH with $\rm SNR\sim 9$ (see Figure~\ref{fig:bayesian1} in the appendix), which validates the parameter measurement accuracy for sources in the marginal detection case.

Using the aforementioned method, we showcase three representative systems from the mock Milky Way GC catalog, with $e = 0.1671 - 0.8969$ (see Table~\ref{tab:compaison} in the appendix). To better describe the sources' localization accuracy, we convert the posterior of their sky location angles, $(\phi,\, \theta)$, into the solid angle of the corresponding sky area $\delta \Omega_{\rm Bayesian}$ ($68\%$ credible region). We then compare $\delta \Omega_{\rm Bayesian}$ with the area of sky error ellipsoid from Fisher matrix analysis, $\delta \Omega_{\rm Fisher}$, and the size of their host GCs. As shown in Table~\ref{tab:compaison}, the Fisher matrix and Bayesian analysis generally yield consistent results in the sources' sky localization accuracy, although the Fisher results are systematically smaller by a factor of $\sim 2$ in angular size estimation (or $\sim 5$ in sky area). Furthermore, for high-SNR BBHs (see, e.g., the three systems with $\rm SNR = 35-94$ in Table~\ref{tab:compaison}), their sky area accuracy from both Fisher matrix and Bayesian analysis, $\sim 0.40, 1.00, 1.85$ deg$^2$, are comparable with (or even smaller than) the size of their host clusters, $\sim 2.84, 2.13, 0.52$ deg$^2$.\footnote{However, marginally detectable BBHs ($\rm SNR \sim 8$) can have poor sky localization. For example, the system in Figure~\ref{fig:bayesian1} has a sky localization error of $\sim $ 233 ~deg$^2$ in Bayesian analysis, far exceeding the angular size of its host cluster NGC7078 ($\sim 0.39$~deg$^2$).} These results further confirm that BBHs in the Milky Way can be localized to the sky area of specific host GCs, especially for significant detection cases.

Notably, Figures~\ref{fig:errorsum}-\ref{fig:bayesian} and Table~\ref{tab:compaison} indicate a distance measurement error of $\delta R\sim 1-100$~kpc for the Milky Way BBH sources, which is much larger than the tidal diameter of globular clusters (typically on the order of a few tens of pc). Therefore, LISA is unlikely to directly associate the spatial position of a BBH system with its host GC because of the large uncertainty in the radial distance measurement. However, when determining the host environment of BBH sources, this uncertainty can be resolved, mostly because stellar density in the Milky Way halo is much lower than that of GCs. In other words, the total number of stars in a BBH-hosting GC ($\sim 10^{5}-10^{6}$, see Figure~\ref{fig:GCs}) is much larger than the total number of halo stars in other parts of the error volume\footnote{Note that most GCs reside in the Galatic halo \citep[see,e.g., ][]{Harris2010}.}, which makes it highly possible that a detected source is hosted in the cluster instead of other regions of the local universe.

For example, assuming a halo stellar mass density of $\rho_{\rm halo}\sim 10^{-4} \rm M_{\odot}/ pc^{3}$ \citep{Deason_2019}, the total halo stellar mass in the error volume $V_{\rm err}$ of GW detection can be estimated as:
\begin{equation}
    M_{\rm halostars}=\rho_{\rm halo} \times V_{\rm err}\sim \rho_{\rm halo} R^2 \delta R \delta\Omega_{\rm Bayesian}\ ,
    \label{eq:mhalo}
\end{equation}
which yields $M_{\rm halostars}\sim 51 \rm M_{\odot},1.5\times 10^{3}\rm M_{\odot},6.5\times 10^{4} \rm M_{\odot}$ for the first three confidently-localized systems in Table~\ref{tab:compaison}. These values are much smaller than the total stellar mass of their host GCs, which is estimated as $M_{\rm GC}\sim 7\times10^5\rm M_{\odot}, 1.5\times 10^6\rm M_{\odot}, 2\times10^5\rm M_{\odot}$ for NGC104, NGC5139, and NGC6341, respectively \citep[see, e.g.,][]{Marks_2010}. 

We note that the relative contributions of GCs and the Galactic field to the total BBH merger rate remain uncertain \citep[see][for a review]{Mandel_2022}. However, recent studies suggest both are likely contributing at a comparable rate \citep[see, e.g.,][]{Zevin+21,Fishbach+23}. Furthermore, even if the galactic field has a larger contribution to the \textit{total} merger rate, considering the significant difference in total stellar mass between the field \citep[$\sim 10^{10}\rm M_{\odot}$ for the Milky Way;][]{Cautun_2020} and GCs \citep[$\sim 10^{7}\rm M_{\odot}$ for the Milky Way;][]{Harris2010}, the BBH merger rate \textit{per unit stellar mass} in GCs is likely comparable to, or higher than in the field. Thus, we assume that all stellar mass (in the Galactic halo and in globular clusters) has an equal probability of hosting a BBH source. Under this assumption, the probability that a given BBH resides in a GC can be estimated as:
\begin{equation}
    P_{\rm incluster}\sim \frac{M_{\rm GC}}{M_{\rm GC}+M_{\rm halostars}}\ .
    \label{eq:p_inhalo}
\end{equation}
Therefore, once a source is accurately located in the sky area of a GC, the probability for the system to be a GC BBH source can be significantly boosted, especially because a globular cluster contributes to the majority of stellar mass in the error volume. Additionally, the highly eccentric formation nature of BBHs in GC (see, e.g., Section~\ref{subsec:detectability}) can also help with distinguishing them from other BBHs from isolated evolution channels in the galactic field, which are expected to have negligible eccentricities \citep[e.g.,][]{Belczynski16,Stevenson17,Eldridge19}. \footnote{However, the localization method discussed earlier in this section is agnostic to the expected eccentricity distributions of different BBH formation channels. For example, if a population of eccentric BBHs is accurately localized toward the Galactic field rather than any GCs, this method can also distinguish it from the in-cluster BBH population.}

\section{Discussion}
\label{sec:discussion}


In this work, we explore the realistic detectability and parameter measurement accuracy of BBHs in Galactic GCs for observations with LISA. Particularly, since GCs are considered ideal environments for the dynamical formation of BBHs \citep[see, e.g.,][]{miller02hamilton,rodriguez15,Samsing+18}, constraining their detectability potential is of prime importance for the community. Notably, \citet{kremer18} predicted that LISA might detect a significant number of BBH sources from $\sim 150$ GCs in the Milky Way, with various orbital parameters. 
Furthermore, many dynamically-formed BBHs can undergo a wide ($a \gtrsim 0.1\,\rm au$), highly eccentric ($e \gtrsim 0.9$) progenitor phase before merger \citep[][]{Kocsis_2012,Hoang+19,Xuan+23b,Xuan24bkg,Xuan24parameter,knee2024detectinggravitationalwaveburstsblack}, which has been recently proposed to have unique imprints on the mHz GW detection. By detecting these eccentric sources, we can distinguish between different formation mechanisms \citep[][]{east13, samsing14, Coughlin_2015,breivik16,vitale16,nishizawa16b,Zevin2017, Gondan_2018a, Gondan_2018b, Lower18, Romero_Shaw_2019, moore19,2021ApJ...913L...7A, 2021arXiv211103634T,Zevin_2021}, enhance the parameter measurement accuracy \citep[such as the orbital frequency evolution and source location, see][]{Xuan23acc,Xuan24parameter}, and probe the potential presence of tertiary companions through eccentricity oscillations \citep{Thompson+11,Antognini+14,Hoang+18,Stephan+19,Martinez+20,Hoang+20,Naoz+20,Stephan+19,Wang+21, Knee2022ecc}.

Using \texttt{CMC Cluster Catalog} of \citet{Kremer_2020} we generate a best-fit model for each Milky Way globular cluster \footnote{Which is computed using the Monte Carlo $N$-body dynamics code \texttt{CMC}.}, and estimate the BBH population within these clusters (illustrated in Figure~\ref{fig:population}). We then generate the BBHs' GW signals using the x-model \citep[][]{Hinder+10}, which incorporates the dynamics of eccentric compact binaries up to 3PN order. Additionally, the waveform analysis includes the detector's annual motion around the Sun so that our results can represent the realistic detection of GW signals. More details of the waveform model and analysis method can be found in \citet{Xuan24parameter}.

Based on the simulation, we compute the number of detectable BBHs formed in Galactic GCs (see Section~\ref{subsec:detectability}). In total, we expect the GW signal from $0.7\pm 0.7$, $2.0\pm 1.7$, $3.6\pm 2.3$, $13.4\pm 4.7$ BBHs to exceed the threshold of $\rm SNR =30$, 5, 3, and 1, respectively, for a 10-year observation of LISA. Among all these sources, in-cluster BBHs contribute to 0.7, 1.5, 2.2, and 5.8 GW sources, and the ejected population contributes to $\sim 0$, 0.5, 1.4, and 7.6 GW sources above the threshold of $\rm SNR =30$, 5, 3, 1. We highlight that $\sim 50\%$ of the BBHs with $\rm SNR>5$ has high eccentricity in the detection ($e\gtrsim0.9$), which reflects the dominant population of eccentric BBHs in GCs (as they are long-living sources, see, e,g., Equation~(\ref{eq:lifetime})), and indicates that most of the GW signals from BBHs in Milky Way GCs are characterized by highly eccentric ``GW bursts" in future LISA detection \citep[see, e.g.,][]{Xuan+23b}.

Furthermore, we analyze the properties of Galactic GCs that are most likely to host resolvable LISA sources (see Figure~\ref{fig:GCs}). Specifically, we calculate the expected probability for a cluster to host a BBH system with $\rm SNR>1$ during a 10-year LISA observation. As shown in Figure~\ref{fig:GCs}, these BBH-hosting GCs tend to cluster within specific regions of the parameter space, where the GCs have a close distance to the detector ($R\lesssim 10$~kpc), exhibit a small $r_c/r_h$ ($\sim 0.1-0.3$), and have a large total mass. 


To evaluate the measurement accuracy achievable with LISA, we performed a Fisher matrix analysis on the simulated BBH population, as illustrated in Figures~\ref{fig:mapfisher2} and \ref{fig:errorsum}. Particularly, Figure~\ref{fig:mapfisher2} maps the intrinsic parameters, $(a, 1-e)$, of the simulated Milky Way GC BBH population and uses the background color to show their different parameter measurement errors (5-year observation); Figure~\ref{fig:errorsum} depicts the measurement error of $\{f_{\rm orb}, e, M, q, R, 2a_s\}$ for all the simulated BBHs with $\rm SNR>3$, across all the 10 realizations, for 5-year observation. As illustrated in both figures, most of the BBHs with $\rm SNR>3$ have well-constrained orbital frequency and eccentricity, reaching $\delta f_{\rm orb}/ f_{\rm orb} \sim 10^{-7}-10^{-5}$ and $\delta e/ e<1$. Furthermore, eccentricity can, in general, enhance the measurement accuracy, with $\delta e/ e$ reaching $\sim 10^{-3}-0.1$ for detectable BBHs with $e>0.7$. On the other hand, BBHs with $\rm SNR>20$ can have a total mass measurement accuracy of $\delta M/M\lesssim 10\%$, but most of the marginally detectable sources may not have a well-constrained total mass and mass ratio. Also, we expect BBHs with $\rm SNR>20$ to be confidently localized with an angular resolution of $\sim 10-100$~arcmins, which enables us to localize these sources in specific Milky Way GCs. 

In Section~\ref{sec:bayesian}, we adopt Bayesian analysis to verify the localization accuracy of BBH sources. As shown in Figure~\ref{fig:bayesian} and Table~\ref{tab:compaison}, the full 10-dimensional Bayesian analysis for individual BBH sources yields parameter measurement error (at the 1$\sigma$ levels of the marginal posterior distributions) that are consistent with those from Fisher matrix analysis, within one order of magnitude. Notably, Table~\ref{tab:compaison} presents localization results for representative systems from the mock Milky Way GC catalog. For BBHs with $\rm SNR=$ 35, 77, and 94, the typical sky localization errors exhibit similar values ($\sim 1$deg$^2$) compared to their host cluster size. This result further confirms LISA’s potential to confidently localize BBHs within individual Galactic GCs. Additionally, Milky Way BBH sources typically have large uncertainty in their radial distance measurement ($\sim 1$–100~kpc). However, as shown in Equations~(\ref{eq:mhalo})-(\ref{eq:p_inhalo}), once a BBH is localized within the sky area of a GC, the probability of its cluster origin is significantly enhanced due to the large number of stars in GCs ($\sim 10^5$–$10^6$ stars) compared to the stars within the rest of the error volume (i.e., the low-density Galactic halo).

Although our collective sample of GC simulations effectively spans the full parameter space of Galactic GCs \citep{Kremer_2020}, this simulation suite is a grid, which inevitably means some specific observed GCs do not necessarily have a strong ``one-to-one'' model match. For example, Ref.~\citet{Rui2021}, 
demonstrated how the \texttt{CMC Catalog} may be augmented with additional models to match particular GCs with observed properties lying in between grid points. In another recent study, \citet{Ye2022_47tuc} showed that especially massive GCs like 47~Tuc (NGC~104) may require additional adjustments (in particular variations to the initial density profile and initial stellar mass function) to produce a precise model match. The ideal solution is to produce a separate model for every single Galactic cluster \citep[for some examples, see][]{Kremer2018,Kremer2019_initialsize,Ye2022_47tuc,Ye2024}; however, this is computationally expensive, 
and outside the scope of the current study. We tested the LISA predictions from our best-fit model for NGC~104 with the predictions from the more precise model in \citet{Ye2022_47tuc}, and found that the number of LISA sources is consistent within a factor of $3$ for SNR$>1$. Furthermore, the multimodality of the posterior, particularly in angular parameters, can pose additional challenges for accurate sky localization \citep[see, e.g.,][]{Marsat_2021}. However, a comprehensive analysis of this effect is beyond the scope of the present work.


Additionally, this work simulates the BBH population in observed GCs of the Milky Way, which does not include other potential GW sources in the local galaxy, such as BBHs formed in the Galactic nucleus, Galactic Field, and evaporated GCs. Thus, the number expectation presented here only serves as a lower bound for the LISA detection. Notably, the high computational cost of $N-$body simulation limits the sample size presented in this paper (in total 10 realizations), which may result in uncertainty of the predicted BBH population. However, the prediction of source number and parameter measurement accuracy in this work is expected to be a realistic estimation, based on one of the most up-to-date simulations of GCs \citep[][]{Kremer_2020}. Also, the population properties presented here agree with previous works with different GC models \citep[see, e.g.,][]{kremer18}, which indicates the robustness of the result that LISA can detect a handful of BBHs formed in the Milky Way GCs.


To conclude, mHz-frequency BBHs in Milky Way GCs have the potential to enable direct test of the role of GCs in the formation of GW sources. Given a 5-10 year LISA observation, these systems typically have highly-resolved orbital frequency ($\delta f_{\rm orb}/ f_{\rm orb} \sim 10^{-7}-10^{-5}$) and eccentricity ($\delta e/ e \sim 10^{-3}-0.1$), as well as a measurable total mass when the signal-to-noise ratio exceeds $\sim20$. Furthermore, these high SNR BBHs can be confidently localized in a specific GC of the Milky Way, with an angular resolution of $\sim 10-100$~arcmins in the sky. Therefore, we highlight the potential of detecting BBHs in Milky Way GCs, which allows for accurate tracking of their long-term orbital evolution, distinguishing the compact binary formation mechanisms, and understanding the properties of GCs in the local galaxy.





\acknowledgments
The authors thank the anonymous referee for their valuable feedback and helpful suggestions. ZX acknowledges partial support from the Bhaumik Institute for Theoretical Physics summer fellowship. ZX and SN acknowledge the partial support from NASA ATP Grant No. 80NSSC20K0505 and from NSF-AST Grant No. 2206428 and thank Howard and Astrid Preston for their generous support. Further, ZX, KK and SN thank LISA Sprint 2024 for organizing an interactive and productive meeting.

\appendix

\begin{figure*}[htbp]
    \centering
    \includegraphics[width=6in]{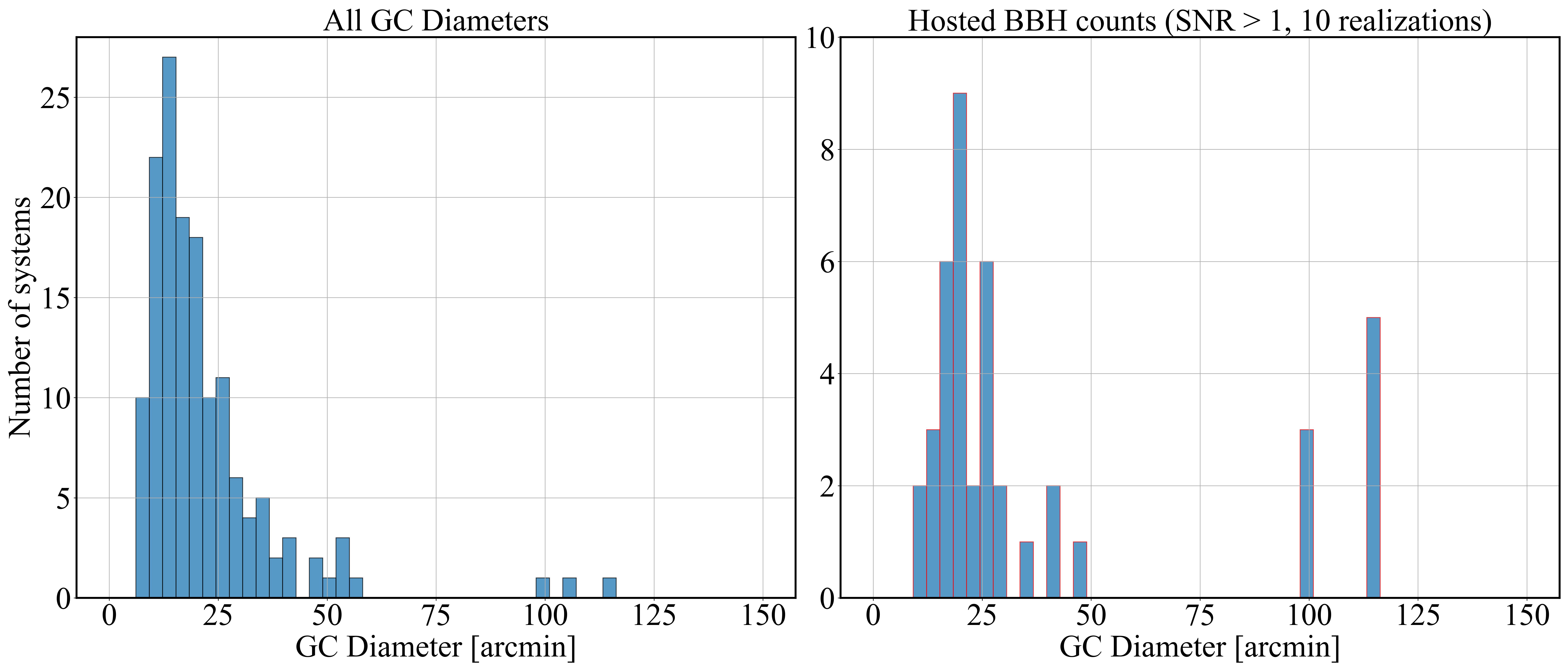}
    \caption{{\bf Histogram of Milky Way GC angular diameters and their hosted BBH counts.} 
    Here we show the distribution of GC angular diameters (left), taken from the Milky Way Star Clusters Catalog \citep{Kharchenko12,Kharchenko13,Schmeja14,Scholz15}, and the number of hosted BBH systems with SNR~$>1$ in each GC diameter bin (right), based on 10 realizations from the \texttt{CMC Cluster Catalog}.
    The difference between the left and right panels arises because detectable BBH systems tend to be hosted by more massive and nearby GCs, which also have a larger angular size.}
    \label{fig:clustersize}
\end{figure*}

\begin{figure*}[htbp]
    \centering
    \includegraphics[width=6in]{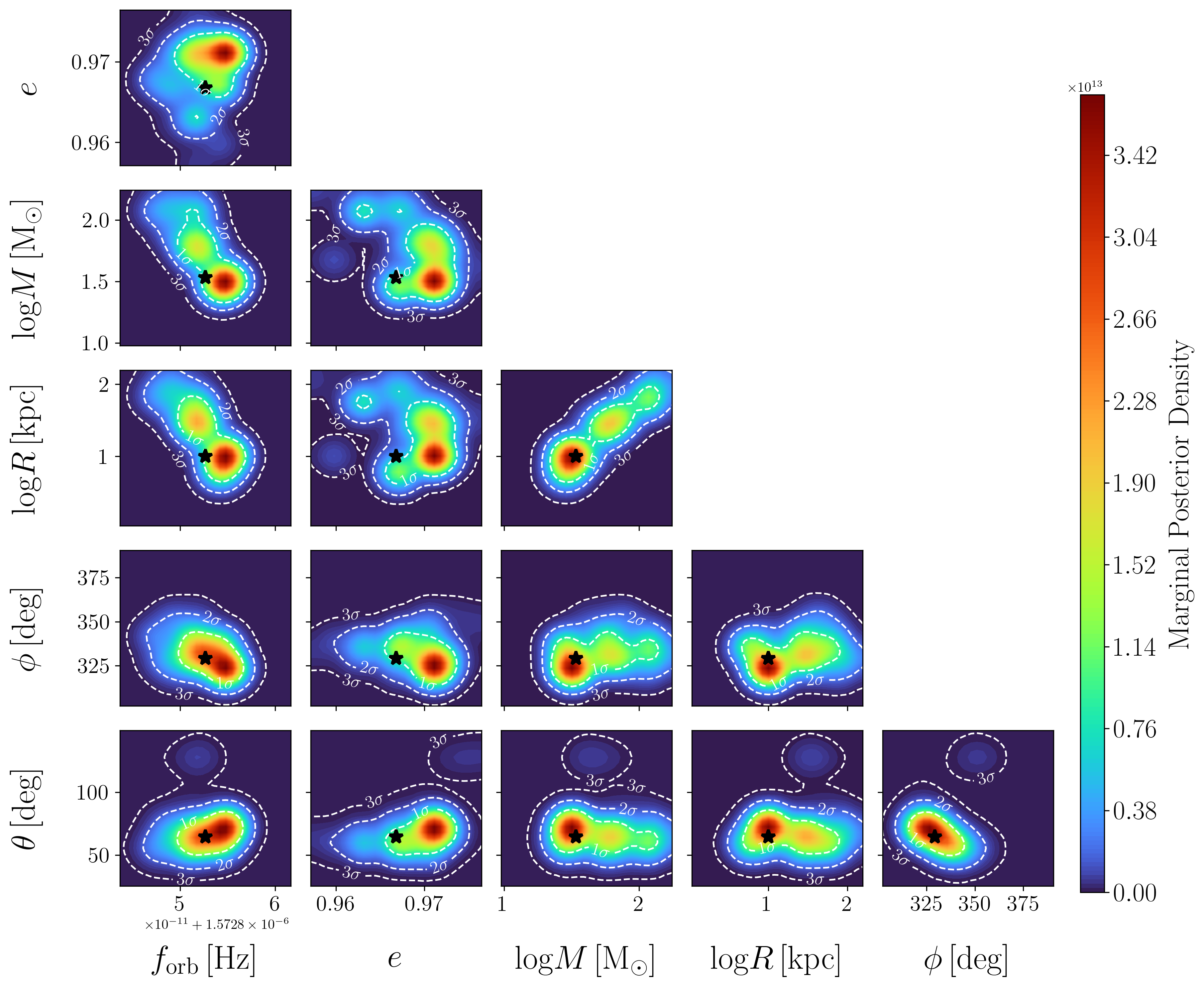}
    \caption{{\bf Parameter Measurement error of an example BBH system in Milky Way Globular Clusters.} Here we adopt the same method as Figure~\ref{fig:bayesian}, but consider a different BBH system with $m_{1}=17.6$~M$_{\odot},$ $m_{2}=16.5$~M$_{\odot},$ $f_{\rm orb}=1.572852\times10^{-6}$~Hz, $e=0.9668,$
    $R=10$~kpc, $\Phi=329.2^{\circ},\,\Theta=64.5^{\circ},\,\phi=341.4^{\circ},\,\theta=149.8^{\circ},\,\psi=14.8^{\circ}$ ($\rm SNR\sim 9$).}
    \label{fig:bayesian1}
\end{figure*}

\begin{deluxetable*}{ccccccccc}
\tablewidth{0pt}
\tablecaption{Sky location and distance measurement accuracy of representative systems in the mock Milky Way GC catalog, compared with their host cluster size (5-yr observation) \label{tab:compaison}}
\tablehead{
\colhead{Host Cluster} & \colhead{Cluster Size} & \colhead{$\delta \Omega_{\rm Bayesian}$} & \colhead{$\delta \Omega_{\rm Fisher}$}& 
 \colhead{$\delta R$} & \colhead{$M$} & \colhead{$a$} & \colhead{$e$} & \colhead{SNR} \\
 \colhead{} & \colhead{[deg$^2$]} & \colhead{ [deg$^2$]} & \colhead{ [deg$^2$]}& 
 \colhead{[kpc]} & \colhead{[$\rm M_{\odot}$]} & \colhead{[au]} & \colhead{} & 
\colhead{(5-yrs)}
}
\startdata
NGC104 & 2.84 & 0.40 & 0.073& 0.25 & 29.5 & 0.01102 & 0.2803 & 94\\
NGC5139 & 2.13 & 1.00 & 0.17 &2.16 &41.9 & 0.06332 & 0.8969 & 77 \\
NGC6341 & 0.52 & 1.85 & 0.38 & 18.0 & 21.2 & 0.00750 & 0.1671 & 35 \\
\enddata
\tablecomments{Here $\delta \Omega_{\rm Bayesian}$ represents the sky area of 68\% credible
regions of the marginal posterior densities (see, e.g., the bottom right panel of Figure~\ref{fig:bayesian}); $\delta \Omega_{\rm Fisher}=\pi a_s b_s$ represents the area of sky error ellipsoid, as estimated by Fisher matrix analysis; $\delta R$ shows the distance measurement accuracy, as of $3\sigma$ level from Bayesian analysis; $M,a,e$ represent the total mass, semi-major axis, and eccentricity of the binary system, respectively. For comparison purposes, we show the name and sky area of each binary's host cluster \citep[inferred from their angular radius in the Milky Way Star Clusters Catalog,][]{Kharchenko12,Kharchenko13,Schmeja14,Scholz15}.
}

\end{deluxetable*}

\bibliography{bibbase}
\end{document}